\documentclass[10pt,a4paper]{article}
\usepackage{jcappub}
\usepackage{pdflscape}
\usepackage{amsmath}
\usepackage{amssymb}
\usepackage{dcolumn}
\usepackage{bm}
\usepackage{color}
\usepackage{epsfig}
\usepackage{amsfonts}
\usepackage{graphicx}
\usepackage{subfigure}
\usepackage{dcolumn}
\begin{document}

\title{Regular black
holes in $f(T)$ Gravity through a nonlinear electrodynamics source}
 
\author[a,e]{Ednaldo L. B. Junior}
\author[a,b]{Manuel E. Rodrigues} 
\author[c,d]{Mahouton J. S. Houndjo}

\affiliation[a]{ Faculdade de F\'{\i}sica, PPGF, Universidade Federal do Par\'{a}, 66075-110, Bel\'{e}m, Par\'{a}, Brazil}

\affiliation[b]{Faculdade de Ci\^{e}ncias Exatas e Tecnologia, Universidade Federal do Par\'{a}\\
Campus Universit\'{a}rio de Abaetetuba, CEP 68440-000, Abaetetuba, Par\'{a}, Brazil}

\affiliation[c]{Institut de Math\'{e}matiques et de Sciences Physiques (IMSP),  
01 BP 613, Porto-Novo, B\'{e}nin}

\affiliation[d]{ Facult\'{e} des Sciences et Techniques de Natitingou - Universit\'{e} de Parakou - B\'{e}nin}

\affiliation[e]{Faculdade de Engenharia da Computa\c{c}\~{a}o, Universidade Federal do Par\'{a}, Campus Universit\'{a}rio de Tucuru\'{\i}, CEP: 68464-000, Tucuru\'{\i}, Par\'{a}, Brazil}

\emailAdd{ednaldobarrosjr@gmail.com}
\emailAdd{esialg@gmail.com}
\emailAdd{sthoundjo@yahoo.fr}


\abstract{We seek to obtain a new class of exact solutions of regular black holes in $f(T)$ Gravity with non-linear electrodynamics material content,  with spherical symmetry in $4D$. The equations of motion provide the regaining of various solutions of General Relativity, as a particular case where the function $f(T)=T$. We developed a powerful method for finding exact solutions, where we get the first new class of regular black holes solutions in the $f(T)$ Theory, where all the geometrics scalars disappear at the origin of the radial coordinate and are finite everywhere, as well as a new class of singular black holes.}


\maketitle


\section{Introduction}
\label{sec1}
General relativity (GR) is formulated in such a way to describe the gravitational effects, that are said, the principle of equivalence, equivalent to the inertial by the Riemannian Differential Geometry. The curvature is a primary concept of this type of geometry, which should be essentially responsible for gravitational effects. The GR has been a theory very effective in describing many local phenomena such as black holes and others which are astro-physical, as well as global phenomena, and the evolution of our universe. But in particular, the description of the evolution of our universe, by the GR, is normally done by the introduction of an exotic fluid, commonly called dark energy, whose the pressure has to be negative. This fluid would be the main content that leads to the accelerated evolution of our current universe. After several models that attempt to describe the Lagrangian associated with the dark energy density, such as the vast majority through the scalar field, there are numerous models that, instead of using the GR for this description, modifies it for this purpose. The first attempts were the use of a Lagrangian density that contained non-linear terms in the curvature scalar  $R$, and so, through the generalization of this idea, there is the best known and addressed modification, which is  $f(R)$ Gravity \cite{fR}. In this theory the action of the system is given by $S_{f(R)}=\int d^4x\sqrt{-g}[f(R)+2\kappa^2\mathcal{L}_{matter}]$, where $f(R)$ is an analytic function of the curvature scalar. Although there were several other modifications, in which the best known are: the Theory $f(R, \mathcal{T})$ \cite{fRT}, with $ \mathcal{T}$ being the trace of the energy-momentum tensor; the $f(\mathcal{G})$ Gravity \cite{fG}, with $\mathcal{G}$ being Gauss-Bonnet term; the $f(R, \mathcal{G})$ \cite{fRG}. All these theories are commonly generalizations of GR in the sense that at some limits of the parameters, they fall in GR.
\par 
Another possible approach to gravitational effects, is one that uses a description of these effects directly by inertia. A geometry also alternative to the Riemannian one, would consider an identically zero curvature, but at the same time a non-null  torsion of the spacetime. Now using the torsion of the space-time as primordial concept to describe the effects of inertia, or equivalently, the gravitational, we can formulate a theory completely equivalent to GR. This is commonly called Weitzenbock spacetime, where the equations of motion are dynamically equivalent of the ones of GR, with a total divergence term more in action. The theory described here is known as Teleparallel Theory (TT) \cite{TT}. We can construct an additional scalar similar to the curvature scalar for the TT, which is represented by $T$, called the torsion scalar. As the equations of TT lead to the same  as that of GR, the problem of accelerated evolution of our universe remains the same. Then there is the possibility of also generalizing the TT, through an action that contains non-linear terms of torsion scalar. The $f(T)$ Gravity \cite{fT} is the generalization of TT, where  $f(T)$ is an analytic function of the torsion scalar.

\par 
Several applications in cosmology were conceived within this $f(T)$ theory, where we can list the main advances in \cite{fT}. In relation to local phenomena such as black holes, there are not yet great advances. The first solution is obtained by Wang in \cite{wang}, with a coupling  with the Maxwell term. Afterword, charged solutions also have been  obtained in $(1 + 2)D$ \cite{gonzalez}. Still in the same way, Capozziello et al. \cite{capozziello1} get charged solutions in $d$-Dimensions. Vacuum Solutions, known in GR, were re-obtained in \cite{ferraro1} with spherical symmetry, and \cite{bejarano} with axial symmetry. The solution of Reissner-Nordstrom-AdS (and their particular cases) was re-obtained in \cite{rodrigues1}, and also with new charged solutions  in $f(T)$ Gravity, where $f(T)$ was reconstructed by a parametric method.
\par 
The initial idea of generalising the TT was the use of an analogy to the action of Born-Infeld \cite{ferraro2}, where there are non-linear terms for the scalar constructed from the Maxwell tensor, $F=(1/4)F^{\mu\nu}F_{\mu\nu}$, where $F^{\mu\nu}$ are the Maxwell tensor components. But the original idea of ​​Born-Infeld \cite{BI} was getting a charged solution where the electric field or by duality the magnetic field, were regular at the origin of the radial coordinate. In the Maxwell's Classical Electromagnetism, the solution to a specific electric punctual charge is divergent where this charge is situated, which for the description with spherical symmetry, is the radial coordinate origin. This was regularised by the Born-Infeld solution. Thus arises the possibility of solutions, coupled with the  Non-Linear Electrodynamics (NED) Lagrangian density, where there is no singularity in spacetime, which has been noticed in the work of Hoffmann and Infeld \cite{hoffmann} still at $1937$, where they found a regular solution. Long time later, Peres obtained another solution where the space-time is free of any type of singularity \cite{peres}. A few years later, Bardeen suggests a magnetic solution where the space-time is completely regular \cite{bardeen}. Thus, new regular solutions emerged \cite{regular} and are still fully explored nowadays \cite{regular2}.
\par 
Exact solutions in GR commonly are obtained containing a singularity, or singular region, which prevents the classification of space-time as globally hyperbolic, thus appearing point or non-physical region for a given measurement singularity. As in the singularities of the space-time geodesics can not be extended, through a finite amount of proper time of a certain measuring apparatus, it disappears, which can not be physically acceptable. For this reason, since the beginning of RG, Einstein could not accept this kind of singular solution. Until now, it seems that this should only be a classic problem, being regularized through quantum effects of gravitation. But one main idea in an attempt to obtain regular exact solutions, is that of Sakharov \cite{sakharov}, where the imposition ``{\it pressure $\equiv$ -energy density}'' (further facilitate obtaining a accelerated current phase of our universe) leads to a breaking at least one energy condition, thus satisfying almost all the singularity theorems for that space-time is always regular. The motivation to use a Lagrangian density of the NED is arising out from the fact of a possible analogy between the effects of polarization of the quantum vacuum, thereby associating the probability of Schwinger formula with the component ($0-0$) for the energy-momentum tensor, thus being facilitated to obtain a new solution that does not impose {\it a priori} the functional form of the Lagrangian density of matter \cite{ansoldi}. 
\par
Recently new charged black hole solutions have been obtained in $f(T)$ theory, coupling  matter content from NED,  demonstrating the possibility of obtaining regular solutions \cite{junior}.  Another exact solution in scalar-torsion gravity  arising of coupling with a scalar field is obtained in \cite{kofinas} for a new wormhole-like solution dressed with a regular scalar field. In this work we focus our attention  on obtaining new regular black hole solutions in the context of $f(T)$ coupled to  NED.
\par
The paper is organized as follows. In section \ref{sec2} the basis of $f(T)$ Gravity coupled with  NED are presented, and the equation of motion. In section \ref{sec3} we show how to reobtain the equations of motion of the cases of regulars black holes, arising from the NED, in  GR. The section \ref{sec4} is devoted to the presentation of the method of solving the equations of motion arising from the subsections  \ref{subsec4.1} for the class of solutions with the fix functional form of $a(r)$, and \ref{subsec4.2} for the class of solutions to the form of a fix $b(r)$.  We make our final considerations in section \ref{sec5}.

\section{The equations of motion in $f(T)$ Gravity}
\label{sec2}
A differentiable manifold has a structure in which the essential elements of a theory of gravitation can be derived from its differential geometry. Through the vectors which are defined in the tangent space to the manifold, and co-vectors, dual elements to vectors, which are defined in the co-tangent space to the manifold, we can then set the line element of the manifold as \begin{eqnarray}
dS^2=g_{\mu\nu}dx^{\mu}dx^{\nu}=\eta_{ab}e^{a}e^{b}\label{ele}\;,\\
e^{a}=e^{a}_{\;\;\mu}dx^{\mu}\;,\;dx^{\mu}=e_{a}^{\;\;\mu}e^{a}\label{the}\;,
\end{eqnarray}
where $g_{\mu\nu}$ is the metric of the space-time, $\eta_{ab}$ the Minkowski metric, $e^{a}$ the tetrads ( that belongs to the co-tangent space) and $e^{a}_{\;\;\mu}$ and their inverses $e_{a}^{\;\;\mu}$, the tetrads matrices satisfying the relations $e^{a}_{\;\;\mu}e_{a}^{\;\;\nu}=\delta^{\nu}_{\mu}$ and  $e^{a}_{\;\;\mu}e_{b}^{\;\;\mu}=\delta^{a}_{b}$. In terms of these matrix, the root of the determinant of the metric is given by $\sqrt{-g}=det[e^{a}_{\;\;\mu}]=e$. Then, one can define a geometry where the curvature should be set equal to zero. Thus, we have to define a connection for which all the components of the Riemann tensor are identically null. This can be done through the Weitzenbock connection. 
\begin{eqnarray}
\Gamma^{\alpha}_{\mu\nu}=e_{a}^{\;\;\alpha}\partial_{\nu}e^{a}_{\;\;\mu}=-e^{a}_{\;\;\mu}\partial_{\nu}e_{a}^{\;\;\alpha}\label{co}\; .
\end{eqnarray}
\par
Now we can define two other important tensors to describe the geometry, that of the torsion and the contorsion 
\begin{eqnarray}
T^{\alpha}_{\;\;\mu\nu}&=&\Gamma^{\alpha}_{\nu\mu}-\Gamma^{\alpha}_{\mu\nu}=e_{i}^{\;\;\alpha}\left(\partial_{\mu} e^{i}_{\;\;\nu}-\partial_{\nu} e^{i}_{\;\;\mu}\right)\label{tor}\;,\\
K^{\mu\nu}_{\;\;\;\;\alpha}&=&-\frac{1}{2}\left(T^{\mu\nu}_{\;\;\;\;\alpha}-T^{\nu\mu}_{\;\;\;\;\alpha}-T_{\alpha}^{\;\;\mu\nu}\right)\label{cont}\; .
\end{eqnarray}
\par
Through these tensors we can construct  a scalar analogous to the scalar curvature of the GR, but it is still best to first define a new tensor, from the torsion and contorsion, to simplify the expression of the torsion scalar. Thus, we define
\begin{eqnarray}
S_{\alpha}^{\;\;\mu\nu}=\frac{1}{2}\left( K_{\;\;\;\;\alpha}^{\mu\nu}+\delta^{\mu}_{\alpha}T^{\beta\nu}_{\;\;\;\;\beta}-\delta^{\nu}_{\alpha}T^{\beta\mu}_{\;\;\;\;\beta}\right)\label{ts0}\;.
\end{eqnarray}
From the components of the torsion scalar and the ones of the torsion  $S_{\alpha}^{\;\;\mu\nu}$, we finally define the torsion scalar as  
\begin{equation}
T=T^{\alpha}_{\;\;\mu\nu}S_{\alpha}^{\;\;\mu\nu}\label{t1}\,.
\end{equation}
We now obtain the equations of motion taking a mater content as a non-linear electrodynamics. For this theory, the Lagrangian density is given by
\begin{eqnarray}
\mathcal{L}=e\left[f(T)+2\kappa^2\mathcal{L}_{NED}(F)\right],\label{lagrangean}
\end{eqnarray} 
where $\mathcal{L}_{NED}(F)$ is the contribution of the non-linear electrodynamics (NED), and the scalar  $F=(1/4)F_{\mu\nu}F^{\mu\nu}$ for $F_{\mu\nu}$ being the Maxwell tensor components, and $\kappa^2=8\pi G/c^4$, where $G$ is the Newton constant and $c$ the speed of light. The field equations for this theory are obtained through the Euler-Lagrange equations. Thus, one has first to take the following derivatives with respect to the tetrads

\begin{eqnarray}
&&\frac{\partial\mathcal{L}}{\partial e^{a}_{\;\;\mu}}=f(T)ee_{a}^{\;\;\mu}+ef_{T}(T)4e^{a}_{\;\;\alpha}T^{\sigma}_{\;\;\nu\alpha}S_{\sigma}^{\;\;\mu\nu}+2\kappa^2\frac{\partial\mathcal{L}_{NED}}{\partial e^{a}_{\;\;\mu}}\,\\
&&\partial_{\alpha}\left[\frac{\partial\mathcal{L}}{\partial (\partial_{\alpha}e^{a}_{\;\;\mu})}\right]=-4f_{T}(T)\partial_{\alpha}\left(ee_{a}^{\;\;\sigma}S_{\sigma}^{\;\;\mu\nu}\right)-4ee_{a}^{\;\;\sigma}S_{\sigma}^{\;\;\mu\gamma}\partial_{\gamma}T f_{TT}(T)+2\kappa^2\partial_{\alpha}\left[\frac{\partial\mathcal{L}_{NED}}{\partial (\partial_{\alpha}e^{a}_{\;\;\mu})}\right]\,,
\end{eqnarray}
where we define $f_{T}(T)=df(T)/dT$ e $f_{TT}(T)=d^2f(T)/dT^2$. Through the above expressions and the ones of  Euler-Lagrange
\begin{eqnarray}
\frac{\partial\mathcal{L}}{\partial e^{a}_{\;\;\mu}}-\partial_{\alpha}\left[\frac{\partial\mathcal{L}}{\partial (\partial_{\alpha}e^{a}_{\;\;\mu})}\right]=0\label{ELeq}\,,
\end{eqnarray} 
and by multiplying all by the factor  $e^{-1}e^{a}_{\;\;\beta}/4$, we get the following equations of motion 
\begin{eqnarray}
S_{\beta}^{\;\;\mu\alpha}\partial_{\alpha}T f_{TT}(T)+\left[e^{-1}e^{a}_{\;\;\beta}\partial_{\alpha}\left(ee_{a}^{\;\;\sigma}S_{\sigma}^{\;\;\mu\alpha}\right)+T^{\sigma}_{\;\;\nu\beta}S_{\sigma}^{\;\;\mu\nu}\right]f_{T}(T)+\frac{1}{4}\delta^{\mu}_{\beta}f(T)=\frac{\kappa^2}{2}\mathcal{T}^{\;\;\mu}_{\beta}\label{meq}
\end{eqnarray} 
where $\mathcal{T}_{\beta}^{\;\;\mu}$ is the energy-momentum tensor of non-linear electrodynamics source 
\begin{eqnarray}
\mathcal{T}_{\beta}^{\;\;\mu}=-\frac{2}{\kappa^2}\left[\delta^{\mu}_{\beta}\mathcal{L}_{NED}(F)-\frac{\partial \mathcal{L}_{NED}(F)}{\partial F}F_{\beta\sigma}F^{\mu\sigma}\right]\label{emt}\,.
\end{eqnarray} 
The choice of the set of tetrads occurs such that all equations are consistent. In our particular case we obtain symmetric and static spherically solutions, so we must choose consistently tetrads with this final order. If we choose a set of tetrads diagonal as $[e^{a}_{\;\;\mu}]=diag[e^{a(r)/2},e^{b(r)/2},r,r\sin\theta]$, we fall back to an inconsistency for the equations of motion. It can be seen in \cite{rodrigues4} that with this choice, an undesired relationship appears between the field equations, for the components  $1-2 (r-\theta)$, which forces the theory to be restricted to TT, where $f(T)$ is linear in $T$. So we make here another choice of tetrads. One possibility, consistent with the equations of motion, is that for which we obtain a set of tetrads through a Lorentz transformation for the diagonal case, leading to \cite{rodrigues5}.

\begin{eqnarray}\label{ndtetrad}
\{e^{a}_{\;\;\mu}\}=\left[\begin{array}{cccc}
e^{a/2}&0&0&0\\
0&e^{b/2}\sin\theta\cos\phi & r\cos\theta\cos\phi &-r\sin\theta\sin\phi\\
0&e^{b/2}\sin\theta\sin\phi &
r\cos\theta\sin\phi &r\sin\theta
\cos\phi  \\
0&e^{b/2}\cos\theta &-r\sin\theta  &0
\end{array}\right]\;,
\end{eqnarray}  
where we define the determinant of the tetrad by  $e=det[e^{a}_{\;\;\mu}]=e^{(a+b)/2}r^2\sin\theta$. With these tetrads we can reconstruct the metric with spherical symmetry 
\hspace{0,2cm}
\begin{equation}
dS^2=e^{a(r)}dt^2-e^{b(r)}dr^2-r^2\left[d\theta^{2}+\sin^{2}\left(\theta\right)d\phi^{2}\right]\label{ltb}\;,
\end{equation}
where the metric parameters $\{a(r),b(r)\}$ are assumed to be functions of radial coordinate $r$ and are not time dependent. 
\par 
Let us calculate the non-zero components of the torsion  (\ref{tor}) and contorsion (\ref{cont}) 
\begin{eqnarray}
T^{0}_{\;\;10}=\frac{a'}{2}\,,\;\;\;T^{2}_{\;\;21}=T^{3}_{\;\;31}=\frac{e^{b/2}-1}{r}\,,\,K_{\;\;\;\;0}^{10}=\frac{a'e^{-b}}{2}\,,\;\;\;K_{\;\;\;\;1}^{22}=K_{\;\;\;\;1}^{33}=\frac{e^{-b}(e^{b/2}-1)}{r}\,\,.\label{tt}
\end{eqnarray}
We also calculate the non-zero components of the tensor $S_{\alpha}^{\;\;\mu\nu}$ in (\ref{ts0}), giving
\begin{eqnarray}
S_{0}^{\;\;01}=\frac{e^{-b}(e^{b/2}-1)}{r}\,,\;\;S_{2}^{\;\;12}=S_{3}^{\;\;13}=\frac{e^{-b}\left(a'r-2e^{b/2}+2\right)}{4r}\,.\label{ts1}
\end{eqnarray}
\par
Through the torsion tensor components (\ref{tt}) and of (\ref{ts1}), we can calculate the torsion scalar (\ref{t1}), which results in 
\begin{equation}
T=\frac{2}{r}\left[-\left(a'+\frac{2}{r}\right)e^{-b/2}+\left(a'+\frac{1}{r}\right)e^{-b}+\frac{1}{r}\right]  \label{te}\,.
\end{equation}
In general, the torsion scalar depends on the radial coordinate, as we will see below. 
\par
Let us define the $4$-potential electric being a 1-form $A=A_{\mu}dx^{\mu}$, where the components are given by  $\{A_{\mu}\}=\{A_{0}(r),0,0,0\}$ and  $A_{0}$ is the electric potential because of the existence of just one charge. Let us define the Maxwell tensor as a 2-form  $F=F_{\mu\nu}dx^{\mu}\wedge dx^{\nu}$, where the components are anti-symmetric and given by $F_{\mu\nu}=\nabla_{\mu}A_{\nu}-\nabla_{\nu}A_{\mu}$. With the specific choice of the spherical symmetry and the metric being static, there exists just an unique component of the Maxwell tensor  $F_{10}(r)$ \cite{plebanski}. The field equations for  $A_{\mu}$, using the Lagrangian density (\ref{lagrangean}), are given by  
\begin{equation}
\nabla _{\mu}\left[F^{\mu\nu}\frac{\partial\mathcal{L}_{NED}}{\partial F}\right]=\partial_{\mu}\left[e^{-1}F^{\mu\nu}\frac{\partial\mathcal{L}_{NED}}{\partial F}\right]=0\label{mMaxeq}\,,
\end{equation}
whose solution, within  $\nu=0$, is given by
\begin{eqnarray}
F^{10}(r)=\frac{q}{r^2}e^{-(a+b)/2}\left(\frac{\partial\mathcal{L}_{NED}}{\partial F}\right)^{-1}\,,\label{F10}
\end{eqnarray}
where $q$ is the electric charge.
\par
Through the above result, the field equations (\ref{meq}), with (\ref{emt}), are given by 
\begin{eqnarray}
&&2\frac{e^{-b}}{r}\left(e^{b/2}-1\right)T'f_{TT}+\frac{e^{-b}}{r^2}\left[b'r+\left(e^{b/2}-1\right)\left(a'r+2\right)\right]f_T+\frac{f}{2}=-2\mathcal{L}_{NED}-2\frac{q^2}{r^4}\left(\frac{\partial\mathcal{L}_{NED}}{\partial F}\right)^{-1}\label{eq1}\,,\\
&&\frac{e^{-b}}{r^2}\left[\left(e^{b/2}-2\right)a'r
+2\left(e^{b/2}-1\right)\right]f_T+\frac{f}{2}=-2\mathcal{L}_{NED}-2\frac{q^2}{r^4}\left(\frac{\partial\mathcal{L}_{NED}}{\partial F}\right)^{-1}\label{eq2}\,,\\
&&\frac{e^{-b}}{2r}\left[a'r+2\left(1-e^{b/2}\right)\right]T'f_{TT}+
\frac{e^{-b}}{4r^2}\Big[\left(a'b'-2a''-a'^2\right)r^2+\nonumber\\
&&+\left(2b'+4a'e^{b/2}-6a'\right)r-
4e^{b}+8e^{b/2}-4\Big]f_{T}+\frac{f}{2}=-2\mathcal{L}_{NED}\label{eq3}\,.
\end{eqnarray}

In the next section we will re-obtain some of the most famous regular black holes solutions in  GR.

\section{Recovering some known solutions in General Relativity}\label{sec3}

In this section we will show that one can re-obtain all the regular black holes solutions in GR. To this end, we must demonstrate that the equations of motion (\ref{eq1})-(\ref{eq3}) fall again to the same differential equations of GR. Then, we will stat by the particular case of the TT, where $f(r)=T(r)$, $f_T(r)=1$ and  $f_{TT}(r)=0$. Hence, by subtracting the equation  (\ref{eq2}) from (\ref{eq1}), one gets the following restriction $a(r)=-b(r)+a_0$, with $a_0\in\Re$. From the symmetry of the temporal translation, one can redefine the temporal coordinate ``$t$'' for obtaining this constant $a_0$. Let us then set $a_0=0$, without loss of generality, such that $b(r)=-a(r)$. As usually done for the case of regular black holes, let us rewrite the metric function  $a(r)$, in terms of another function depending on the radial coordinate as follows
\begin{eqnarray}
a(r)=-b(r)=ln\left[1-\frac{2M(r)}{r}\right]\label{ar1}\,.
\end{eqnarray}
Now, by substituting all this into the field equations  (\ref{eq1})-(\ref{eq3}), one gets 
\begin{eqnarray}
&&\mathcal{L}_{NED}+\frac{q^2}{r^4}\left(\frac{\partial\mathcal{L}_{NED}}{\partial F}\right)^{-1}+\frac{M'(r)}{r^2}=0\label{eq1.1}\,,\\
&&2\mathcal{L}_{NED}+\frac{M''(r)}{r}=0\label{eq3.1}\,.
\end{eqnarray}
These equations are exactly the same as that of GR, coupled with the  NED for regular black holes (see the equations  (7) and  (8) of  \cite{balart}\footnote{Note that we use a definition of energy-momentum tensor in (\ref{emt}) different from the one used by Balart and Vagenas in the equation (6) of  \cite{balart}.}). Thus, all the solutions obtained in GR can be re-obtained through these equations here, as for example taking : $M(r)=m\,e^{-q^2/(2mr)}$, which we will use later, nevertheless, in the context of the non-linear  $f(T)$ theory.


\section{Analogy between the known solutions and new  black hole solutions}\label{sec4}
The differential equations of motion for the $f(T)$ theory coupled with the NED source, are differential non-linear coupled equations, rendering difficult how to obtain exact solution. But here, we will present a powerful algorithm for obtaining exact solutions for this theory. The basic idea is to use the Lagrangian density of NED, $\mathcal{L}_{NED}$, and it derivative with respect to the scalar $F$, $\partial \mathcal{L}_{NED}/\partial F=\mathcal{L}_F$, as functions that satisfying the field equations. We can impose that  $\mathcal{L}_{NED}$ satisfy (\ref{eq1}), which by symmetry, and others considerations to be seen later, also satisfies (\ref{eq2}). And finally,  $\mathcal{L}_F$ has to satisfy the ultimate equation of motion  (\ref{eq3}). When the unique field component is given by (\ref{F10}), we will see that all the equations of motion are satisfied. But the expressions of $\mathcal{L}_{NED}$ and $\mathcal{L}_F$, satisfying the set of equations  (\ref{eq1})-(\ref{eq3}), should not be related, one as derivative  with respect to $F$ of the other. In such a situation, we impose this relation in order to obtain a consistent solution with this theory, leading to an equation with constraint, which has to be necessarily satisfied.
\par  
Now we will present the algorithm. By subtracting the equation  (\ref{eq2}) from  (\ref{eq1}) one gets the following differential equation 
\begin{eqnarray}
\frac{d}{dr}\left[ln\, f_{T}(T)\right]+\left[\frac{a'+b'}{2(e^{b/2}-1)}\right]=0\label{diff}\,.
\end{eqnarray}
By integrating this equation one gets 
\begin{eqnarray}
f_{T}(r)=\exp\left\{-\frac{1}{2}\int\frac{a'(r)+b'(r)}{[e^{b(r)/2}-1]}dr\right\}\label{fT}\,.
\end{eqnarray}

Note here that when we choose the relation  $b(r)=-a(r)$, from (\ref{fT}), one gets $f_T(r)=1$, which, by integration on  $T$, leads to  $f(T)=T+T_0$, where $T_0\in\Re$, known as the TT theory with cosmological constant, conform to the no-go theorem of \cite{junior}. Then, in order to do not recover the TT, we always have to ensure that  $b(r)\neq -a(r)$. Thus, we have here two possibilities to use the algorithm:  a) fixing a functional form to the metric  $a(r)$, which is usual for regular solutions; b) fixing the functional form for the metric  $b(r)$. We will do it the coming section.

\subsection{Fixing $a(r)$}\label{subsec4.1}

In order to obtain new regular black holes solutions, it is common to rewrite the metric function $a(r)$  as 
\begin{eqnarray} 
a(r)=\ln\left[1-\frac{2M(r)}{r}\right]\label{a}\,,
\end{eqnarray}
where $M(r)$ is an analytic function of the radial coordinate $r$, which falls to the black hole mass in the limit of spacial infinity, regarding asymptotically flat solution. Now we set (\ref{fT}) and  (\ref{a}) in the equation (\ref{eq1}), for obtaining  $\mathcal{L}_{NED}$, 
\begin{eqnarray}
\mathcal{L}_{NED}&=&\frac{1}{2}\Bigg\{-\frac{f(r)}{2}-2\frac{q^2}{r^4\mathcal{L}_{F}}+
\Bigg[\Bigg(2e^{-b(r)}f_T(r)\nonumber\\
&&\Bigg( e^{b(r)/2}M(r)+r\Big(1-e^{b(r)/2}+\left(e^{b(r)/2} \right)M'(r)\Big)\Bigg)\Bigg)\Big/\left(r^2\left(r-2M(r)\right)\right)\Bigg]\Bigg\}\label{L}\,,
\end{eqnarray}
where $\mathcal{L}_{F}=\partial\mathcal{L}_{NED}/\partial F$ and  $M'(r)=dM(r)/dr$. By subtracting (\ref{eq3}) from (\ref{eq1}), one gets $\mathcal{L}_{F}$
\begin{eqnarray}
\mathcal{L}_{F}&=&-\Bigg[4q^2e^{-b(r)}f_T(r)\left(e^{b(r)/2}-1\right)\left(r-2M(r)\right)^2\Bigg]\Bigg/\Bigg[r^2\Big(2r^2-2e^{b(r)/2}r^2-2e^{b(r)}r^2\nonumber\\
&&+2e^{3b(r)/2}r^2+8e^{b(r)}rM(r)-8e^{3b(r)/2}rM(r)-4M^2(r)+6e^{b(r)/2}M^2(r)-8e^{b(r)}M^2(r)\nonumber\\
&&+8e^{3b(r)/2}M^2(r)+2r^3b'(r)-2e^{b(r)/2}r^3b'(r)-6r^2M(r)b'(r)+7e^{b(r)/2}r^2M(r)b'(r)\nonumber\\
&&+4rM^2(r)b'(r)-6e^{b(r)/2}M^2(r)b'(r)-8r^2M'(r)+8e^{b(r)/2}r^2M'(r)+8rM(r)M'(r)-12e^{b(r)/2}rM(r)M'(r)\nonumber\\
&&-2r^3b'(r)M'(r)+e^{b(r)/2}r^3b'(r)M'(r)+4r^2M(r)b'(r)M'(r)-2e^{b(r)/2}r^2M(r)b'(r)M(r)+4r^2(M'(r))^2\nonumber\\
&&-2e^{b(r)/2}r^2(M'(r))^2+2r^3M''(r)-2e^{b(r)/2}r^3M''(r)-4r^2M(r)M''(r)+4e^{b(r)/2}r^2M(r)M''(r)\Big)\Bigg]\label{LF}\,. 
\end{eqnarray}

Now,, with the expressions  (\ref{fT}), (\ref{a}), (\ref{L}) and  (\ref{LF}) the set of the differential equations  (\ref{eq1})-(\ref{eq3}) are satisfied.  We can obtain the non-null components of the Maxwell tensor, and taking  (\ref{LF}) in   (\ref{F10}) one gets
\begin{eqnarray}
F^{10}(r)&=&e^{-3b(r)/2}f_{T}(r)\sqrt{r}\Bigg\{M^2(r)\left[4-6e^{b(r)/2}+8e^{b(r)}-8e^{3b(r)/2}+2(3e^{b(r)/2}-2)rb'(r)\right]\nonumber\\
&&-r^2\Big[rb'(r)\left(2-2e^{b(r)/2}+(e^{b(r)/2}-2)M'(r)\right)+2\Big(4(e^{b(r)/2}-1)M'(r)-(e^{b(r)/2}-2)M'^2(r)\nonumber\\
&&+(e^{b(r)/2}-1)(-1+e^{b(r)}-rM''(r))\Big)\Big]+rM(r)\Big[rb'(r)\left(6-7e^{b(r)/2}+2(e^{b(r)/2}-2)M'(r)\right)\nonumber\\
&&+4\left((3e^{b(r)/2}-2)M'(r)+(e^{b(r)/2}-1)(2e^{b(r)}-rM'(r))\right)\Big]\Bigg\}\Bigg/\left[4q(e^{b(r)/2}-1)(r-2M(r))^{5/2}\right]\label{F10-1}\,.
\end{eqnarray}

Now, we have to impose the relation between the Lagrangian density and its derivative in order to get a consistent solution. We then have the following consistent relation 
\begin{eqnarray}
\mathcal{L}_{F}(r)=\frac{d\mathcal{L}_{NED}}{dr}\left(\frac{dF}{dr}\right)^{-1}\label{imposition},
\end{eqnarray}
where $F=(1/4)F^{\mu\nu}F_{\mu\nu}$. The imposition (\ref{imposition}), within (\ref{L})-(\ref{F10-1}), yields the following differential equation 
\begin{eqnarray}
\left[M(r)\left(2-2rb'(r)\right)+r\left(rb'(r)-2M'(r)\right)\right]\left[\left(1-2e^{b(r)/2}\right)M(r)+r\left(e^{b(r)/2}+M'(r)-1\right)\right]=0\label{eqM1}\,.
\end{eqnarray}
The equation (\ref{eqM1}) possesses two solutions 
\begin{eqnarray}
b(r)=-a(r)\label{b1}\,,
\end{eqnarray}
and still, 
\begin{eqnarray}
b(r)=2 \ln\left\{e^{-a(r)}\left[M'(r)-\frac{M(r)}{r}\right]-1\right\}\label{b2}\,,
\end{eqnarray}
with $a(r)$ given in  (\ref{a}).

Note now that there exist only two solutions which make the solutions of the differential equations being consistent, and satisfying (\ref{imposition}). The first of these solutions, given by (\ref{b1}), falls into, where $f_T(T)=1$ in (\ref{fT}), because $b(r)=-a(r)$, and then does not give any interesting thing, being just an analogous solution to that of GR. Let us now focus out attention to the second solution given in (\ref{b2}). 

\begin{figure}[h]
\centering
\includegraphics[height=6cm,width=10cm]{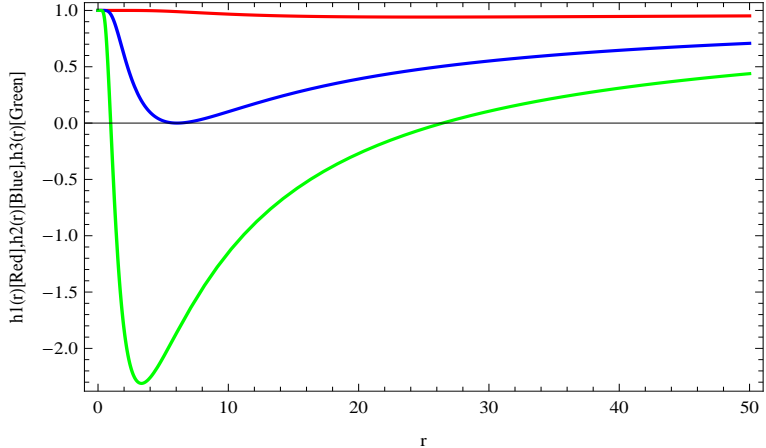}
\caption{\scriptsize{Graphical representation of the function $\exp[-b(r)]$ for the cases $|q|\sqrt{e}/2<m$ (green), $|q|\sqrt{e}/2=m$ (blue) and  $|q|\sqrt{e}/2>m$ (red).}} 
\label{fig1}
\end{figure}

We can now specify some models for $M(r)$, which is commonly done for regular black holes in GR \cite{balart}. Taking a simple model $M(r)=me^{-q^2/(2mr)}$, with $m\in\Re_+$, we can see that the black holes horizon is given by  $g^{11}(r_H)=e^{-b(r_H)}=0$, with $r_H=-[q^2/(2m)]W[-q^2/(2m)^2]$, where $W[y]$ is the Lambert function  \cite{dence}, which solves $W[y]e^{W[y]}=x$. We have a non-degenerate horizon for $|q|\sqrt{e}/2<m$, then $g^{11}(r)=0$ has two real and different roots; for $m=|q|\sqrt{e}/2$ we have a degenerate horizon, and $g^{11}(r)=0$ has a single real root; finally we have no horizon in the case $|q|\sqrt{e}/2>m$ and $g^{11}(r)=0$ has no real root. we represent $g^{11}(r)$ in figure \ref{fig1}.

We still see that this is an asymptotically minkowskian solution, because in the limit of infinite space one gets $e^{a(r)}=e^{b(r)}=1$. The torsion scalar $T(r)$ can be obtained from (\ref{te}), yielding
\begin{eqnarray}
T(r)&=&\frac{6q^4+8q^2r\left(5m-4e^{q^2/(2mr)}r\right)+8r^2\left(3m^2-8e^{q^2/(2mr)}mr+4e^{q^2/mr}r^2\right)}{r^2\left[q^2+2r\left(m-e^{q^2/(2mr)}r\right)\right]^2}\label{te1}\,.
\end{eqnarray}
The torsion scalar (\ref{te1}) goes to zero in the limits $r\rightarrow 0$ and  $r\rightarrow +\infty$. This shows that the torsion scalar is regular at the origem of the radial coordinate, also indicating a regular solution.
\par 
As in some cases of charged solution, the torsion scalar can be regular, but it can exist non-detected singularity by this method   \cite{capozziello1}, and we will calculate the curvature scalar and other ones which exhibit singularity in GR. We can calculate the curvature scalar  $R=g^{\mu\alpha}g^{\nu\beta}R_{\mu\nu\alpha\beta}$, where $R_{\mu\nu\alpha\beta}$ are the components of  Riemann tensor  calculated from the Levi-Civita connection\footnote{{\bf The Levi-Civita connection $\bar{\Gamma}^{\alpha}_{\;\;\mu\nu}$ is symmetric on $(\mu\nu)$ indexes. The Riemann tensor calculated from the Weitzenbock connection $\Gamma^{\alpha}_{\;\;\mu\nu}$ in (\ref{co}) vanishes identically \cite{li,sotiriou}.}} \cite{li,sotiriou}, by the expression $R=-T-2\nabla^{\nu}T^{\mu}_{\;\;\nu\mu}$ \cite{li}. Thus, with the components  (\ref{tt}) and the torsion scalar (\ref{te1}), we get the curvature scalar given by 
\begin{eqnarray}
R(r)=\frac{4\left(2m-e^{q^2/(2mr)}r\right)\left[4m^2q^2r+e^{q^2/(2mr)}q^4+4m^3r^2-mq^2(q^2+4e^{q^2/(2mr)}r^2)\right]}{mr\left[q^2+2r(m-e^{q^2/(2mr)}r)\right]^3}\label{R1}\,.
\end{eqnarray}
The curvature scalar $R(r)$ in (\ref{R1}) converges to zero for two limits $r\rightarrow 0$ and  $r\rightarrow +\infty$, indicating that this solution is regular in all the spacetime. This can be confirmed by the Kretschmann scalar $R^{\alpha\beta\mu\nu}R_{\alpha\beta\mu\nu}$  calculated also for the Levi-Civita connection. This scalar converge to zero for the two limits  $r\rightarrow 0$ and $r\rightarrow +\infty$. 
\par 
Here there is a huge algebraic subtlety to this class of solutions. First we note that the denominator of (\ref{te1}) and (\ref{R1}), as well as for the scalar  $R^{\mu\nu\alpha\beta}R_{\mu\nu\alpha\beta}$, vanishes  with the same roots. So although the scalar being regular in the regions commonly analyzed, infinite space and origin, they diverge in interiot regions to the event horizon. We may check this by placing the curvature scalar, one can be made the same for the other scalars, only in terms of the function $M(r)$
\begin{eqnarray}
R(r)&=&-\frac{2[r-2M(r)]\left\{M(r)^2-2rM(r)[M'(r)+rM''(r)]+r^2[M'(r)^2+rM''(r)]\right\}}{r^2\left\{M(r)+r[M'(r)-1]\right\}^3}\label{Rg1}
\end{eqnarray}
with $M(r)$ given specifically for each model, so it is true for this class of solutions in general. We can see in (\ref{Rg1}), this also occurs on (\ref{R1}), the denominator vanishes to a certain finite value $0<r_s<r_H$. This shows us that there is a singularity in the region $r=r_s$, which is located inside the event horizon, but before the origin. So even if all scalar are regular at the origin, this singular region (hipper-surface), prevents us from accessing it. The causal structure is then being identical to the Schwarzschild black hole in GR, showing that in fact this class solutions are black holes with a singular region covered by the event horizon. This same behavior occurs regardless of the model chosen for $M(r)$.   
\par
\begin{figure}[h]
\centering
\begin{tabular}{rl}
\includegraphics[height=5cm,width=7.5cm]{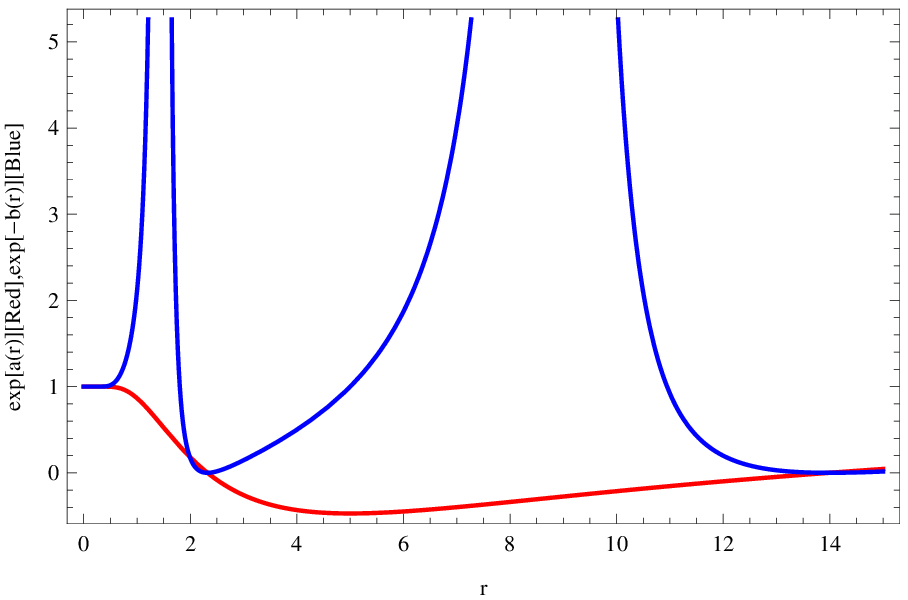}&\includegraphics[height=5cm,width=7.5cm]{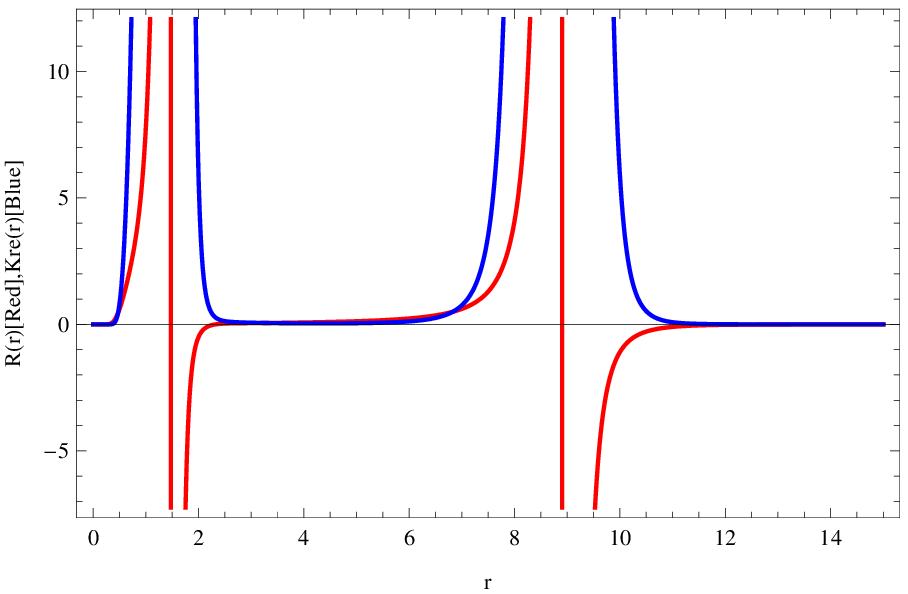}
\end{tabular}
\caption{\scriptsize{Graphical representation of functions $\exp[a(r)],\exp[-b(r)]$ (left), and the functions $R(r),Kre(r)$ (right), in the non-degenerated case, for the values $\{m=q=10\}$.}} 
\label{fig2}
\end{figure}

\begin{figure}[h]
\centering
\begin{tabular}{rl}
\includegraphics[height=5cm,width=7.5cm]{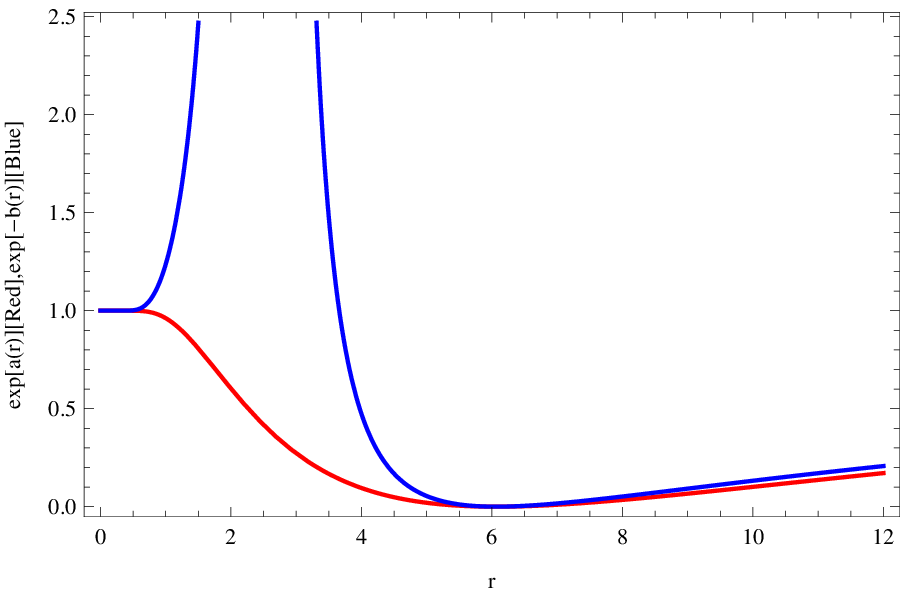}&\includegraphics[height=5cm,width=7.5cm]{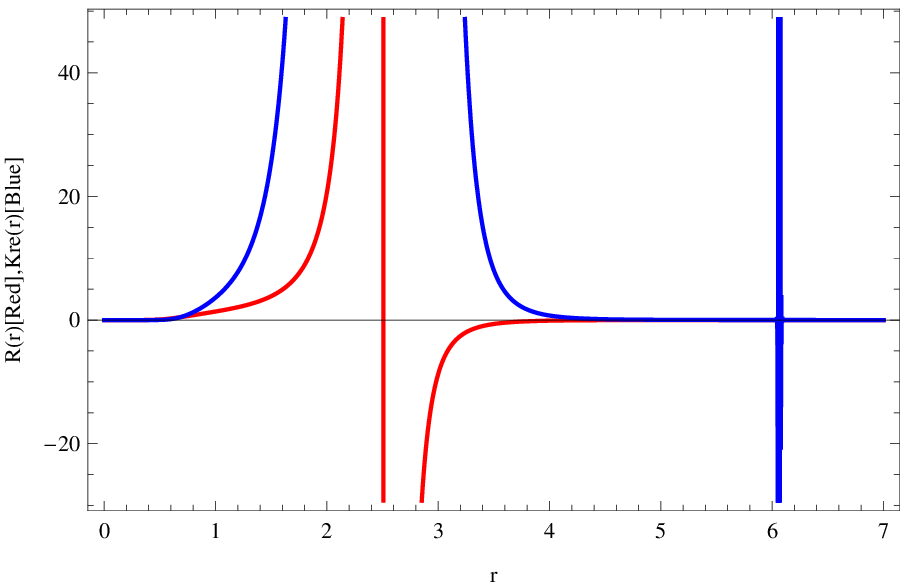}
\end{tabular}
\caption{\scriptsize{Graphical representation of functions $\exp[a(r)],\exp[-b(r)]$ (left), and the functions $R(r),Kre(r)$ (right), in the degenerated case, for the values $\{m=\sqrt{e}q/2,q=10\}$.}} 
\label{fig3}
\end{figure}

 
For this first class of solutions, the metrics functions (\ref{a}) and (\ref{b2}) with $M(r)=me^{-q^2/(2mr)}$, we represent functions $\exp[a(r)],\exp[-b(r)]$, the curvature scalar $R(r)$ and the Kretschmann scalar ($R_{\alpha\beta\mu\nu}R^{\alpha\beta\mu\nu}$) in Figure \ref{fig2}, for the non-degenerated case. We do the same in the figure \ref{fig3} for the degenerated case (extrem case).
We note that although we have the restriction $b(r)\neq -a(r)$, the zeros for  $\exp[a(r)]$ and $\exp[-b(r)]$ still coincides, but the curvature scalar diverges in a value less than the event horizon, characterizing a black hole. In Figure \ref{fig2} we have the zeros of $\{\exp[a(r)],\exp[-b(r)]\}$ on $\{r_{-}=2.322025582334978,r_H=13.98981145146388\}$, and the divergences (zeros for $\{R^{-1}(r),Kre^{-1}(r)\}$) on $\{r_{s1}=1.4719210245449492,r_{s2}=8.906399770461864\}$. Then we found that the actual causal structure is formed by an outer region of the event horizon $r>r_H$, horizon  himself $r=r_H$, and other interior region $r_{s2}<r<r_H$ with a singularity on $r=r_{s2}$. All region $r<r_{s2}$ is excluded due to the impossibility extension of space-time. These are new class of black holes in solutions to $f(T)$ Theory.
\par 
The extreme case also results in new singular solutions. In figure \ref{fig3} we see that there is a single zero to the functions $\{\exp[a(r)],\exp[-b(r)]\}$ on $r_{eH}=6.06530645576941$, and a divergence in $r_{s3}=2.505407627397168$. There are a very high value, the order of $10^{30}$ to the Kretschmann scalar on horizon, but in the singularity, on $r=r_{s3}$, the order is $10^{41}$, what appears to us that this is a solution with regular horizon.   
\par
\begin{figure}[h]
\centering
\begin{tabular}{rl}
\includegraphics[height=5cm,width=7.5cm]{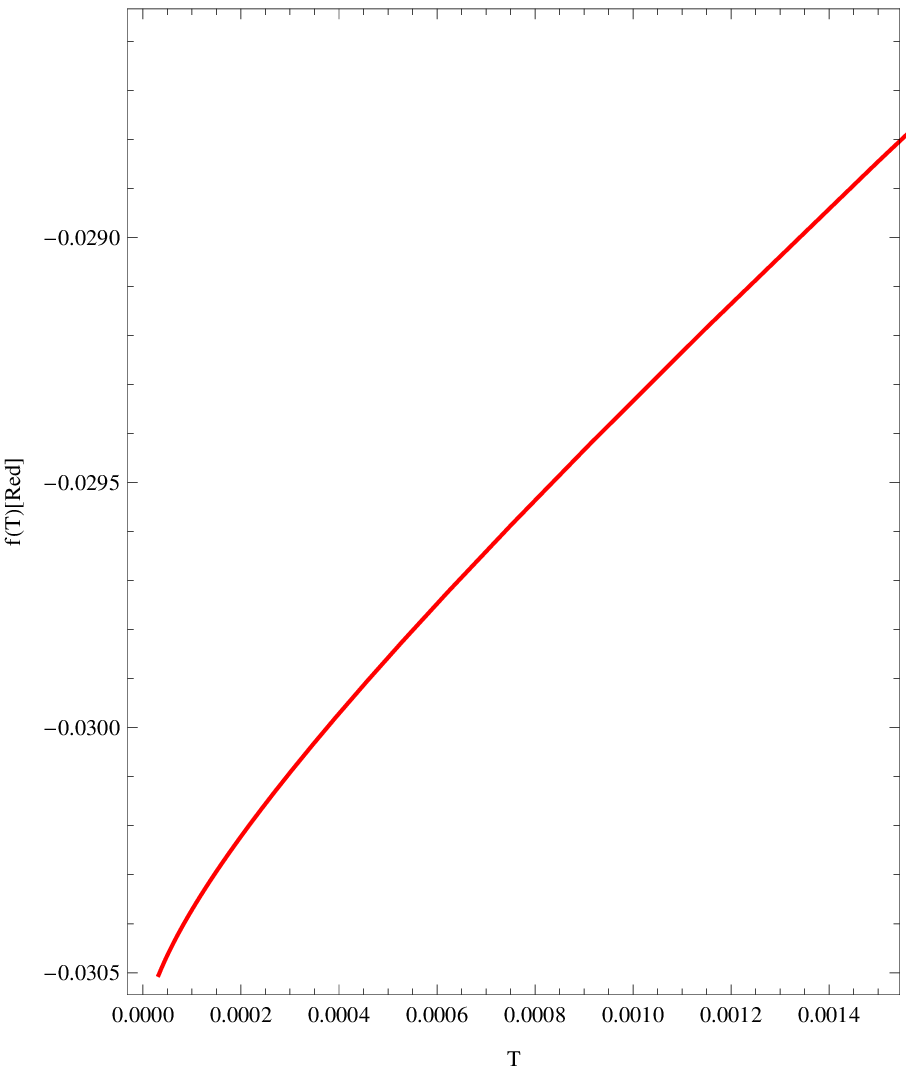}&
\includegraphics[height=5cm,width=7.5cm]{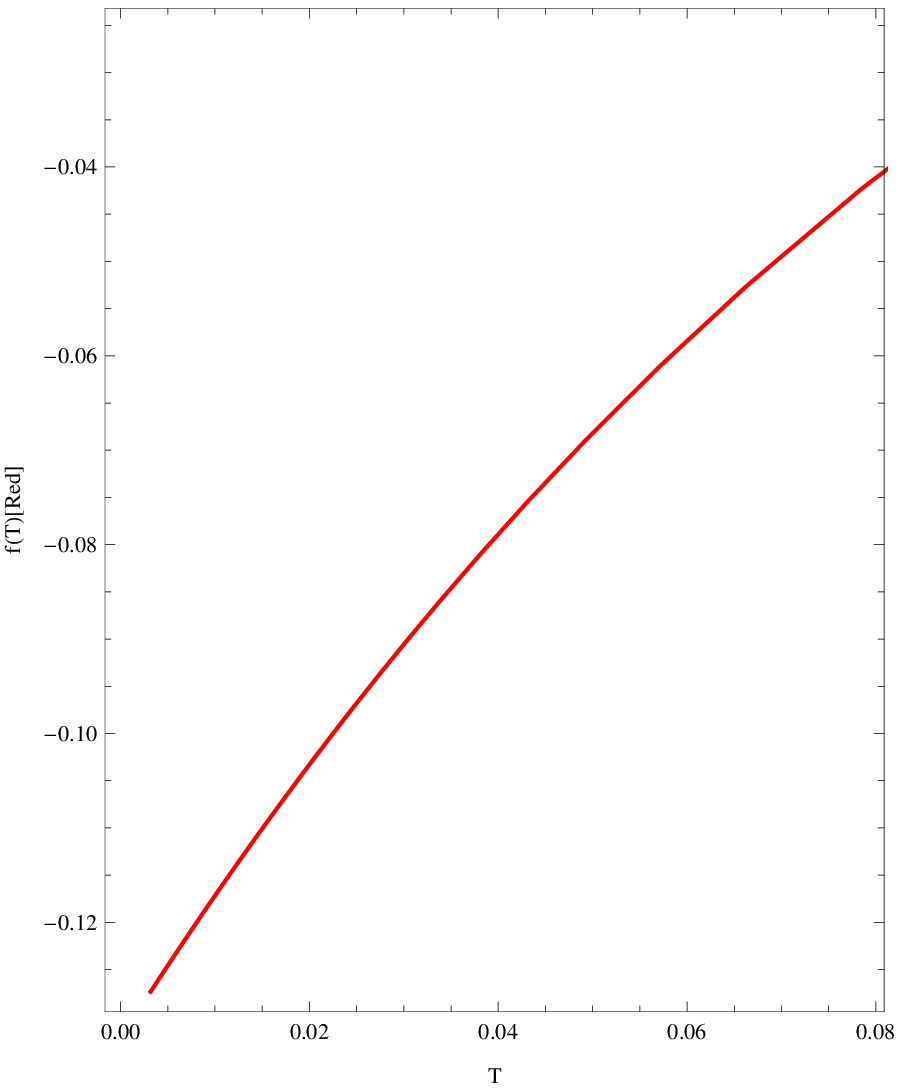}
\end{tabular}
\caption{\scriptsize{Parametric representation to $f(T)\times T$, in the non-degenerated case (left, $r\in[14,500]$) for the values $\{m=q=10\}$, and degenerated case (right, $r\in[6,50]$), for the values $\{m=\sqrt{e}q/2,q=10\}$.}} 
\label{fig3-1}
\end{figure}

 
We calculate numerically the function $f_T(r)$ on (\ref{fT}), using (\ref{a}) and (\ref{b2}) for this model, then we use the chain rule $df/dT=(df/dr)(dr/dT)$ (integrating to determine $f(r)=\int f_T (dT/dr)dr$) obtain at the end the function $f(r)$, which we can represent using a parametric graph in Figure \ref{fig3-1}. We can see in Figure \ref{fig3-1} that the function $f(T)$ does not depend linearly on $T$, then we are not address the TT here.
 
\par 
The conclusion of this section is that the same behavior for this class of solutions can be shown in general, for any model for $M(r)$, as listed in Table $1$ in \cite{balart}. This shows that make a direct analogy between regular solutions to GR $f(T)$ Gravity, results in singular black holes solutions here. This is due to strong restriction $b(r)\neq -a(r)$ arising from the No-Go Theorem \cite{junior}, modifying the expression of the curvature scalar of the GR, as we see in \ref{Rg1}.

\subsection{Fixing $b(r)$}\label{subsec4.2}

We can now perform the same process of the previous section, but now, instead of fixing the functional form of $a(r)$, we directly fix the functional form of $b(r)$. This should be done with  $b(r)=-\ln[1-(2M(r)/r)]$, but this leads to a complicated differential equation. Therefore, we will take into account the form 
\begin{eqnarray}
b(r)=\ln\left[1-\frac{2M(r)}{r}\right]\label{b3}\,.
\end{eqnarray}

Now we can in the same way as before, and in order to satisfy (\ref{eq1}), with (\ref{fT}), we obtain $\mathcal{L}_{NED}(r)$. Afterword we get $\mathcal{L}_F(r)$ in order to satisfy  (\ref{eq3}). Thus, with (\ref{F10}), we get the component  $F^{10}(r)$, from which, using the imposition (\ref{imposition}), leads to the following differential equation 
\begin{eqnarray}
\left[4M(r)+r\left(\sqrt{1-\frac{2M(r)}{r}}(2+ra')-2\right)\right]\left[2M(r)(1-ra')+r(ra'-2M'(r))\right]=0\label{eqM2}\,.
\end{eqnarray}
Then, we have two solutions for (\ref{eqM2}), where the first is  $a(r)=-b(r)$, which, from (\ref{fT}), leads to TT, with $f_T(T)=1$. This first solution is not interesting here, because of its analogy with the GR. The second solution of (\ref{eqM2}) is given by 
\begin{eqnarray}
a(r)=-2\int \frac{1}{r}\left[e^{b(r)/2}-1\right]dr\label{a1}\,.
\end{eqnarray}

As we have to obtain regular black hole solution, the horizon is defined by  $g^{11}(r_H)=e^{-b(r_H)}=0$, which is well described for 
\begin{eqnarray}
M(r)=-\frac{rm(r)}{r-2m(r)}\label{M3}\,,
\end{eqnarray}
because, for this, one has $e^{-b(r)}=1-[2m(r)/r]$. 
\par 
First we analyse in general the scalar of Kretschmann for this class solutions
\begin{eqnarray}
Kre(r)&&=R_{\alpha\beta\mu\nu}R^{\alpha\beta\mu\nu}=-\frac{4}{r^6}\Big\{\left[39-79\sqrt{\frac{r}{r-2m(r)}}\right]m^2(r)+2rm(r)\Big[-30+41\sqrt{\frac{r}{r-2m(r)}}\nonumber\\
&&+\left(-7+6\sqrt{\frac{r}{r-2m(r)}}\right)m'(r)\Big]+r^2\Big[-22\left(-1+\sqrt{\frac{r}{r-2m(r)}}\right)\nonumber\\
&&+3m'(r)\left(m'(r)+2-2\sqrt{\frac{r}{r-2m(r)}}\right)\Big]\Big\}\,.\label{kreg2}
\end{eqnarray}
We note that solutions can only be analytic throughout the space if the restriction $r-2m(r)\geq 0$ is satisfied. As cases of non-degenerate horizons break this condition, we will focus our attention only to the extreme case where this condition is valid. 
\par 
Let us take again one model for  $m(r)$. The model for $m(r)$ is given by $m(r)=m_0e^{-q^2/(2m_0r)}$, with $m_0\in\Re_+$. The events horizon of this solution is given by  $r_H=-[q^2/(2m_0)]W[-q^2/(2m_0)^2]$, with $m_0=q\sqrt{e}/2$. This solution is asymptotically flat and regular in the spacial infinity. The curvature scalar is given by 
\begin{eqnarray}
R(r)=-\frac{e^{-q^2/(2m_0r)}}{r^4}\left\{q^2+2r\left[m_0-e^{q^2/(2m_0r)}r+\sqrt{e^{q^2/(2m_0r)}r\left(-2m_0+e^{q^2/(2m_0r)}r\right)}\right]\right\}\label{R3}\,.
\end{eqnarray}
The curvature scalar and $R^{\mu\nu\alpha\beta}R_{\mu\nu\alpha\beta}$ tend to zero in the limit $r\rightarrow 0$, showing that the solution is regular at the origin of the radial coordinate. 
\par
\begin{figure}[h]
\centering
\begin{tabular}{rl}
\includegraphics[height=5cm,width=7.5cm]{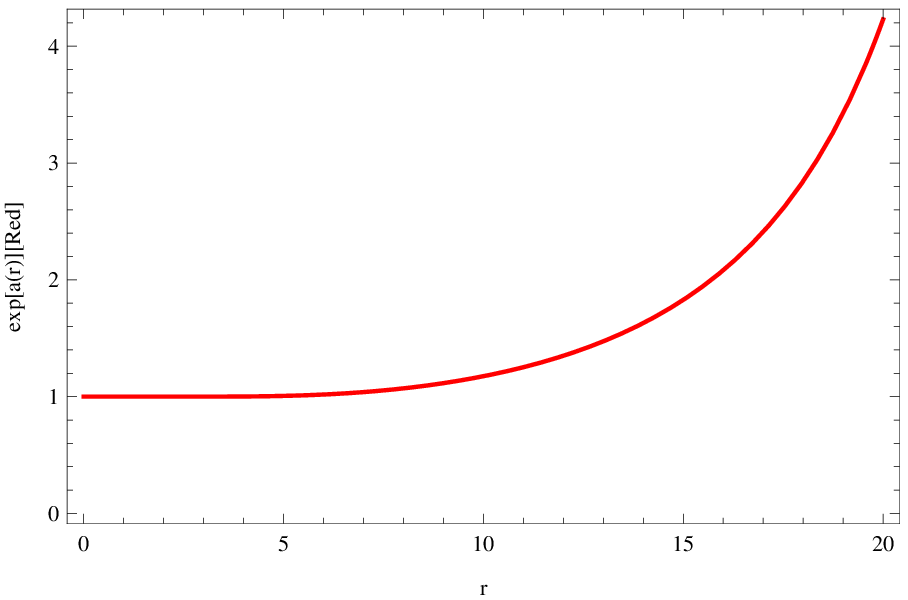}&\includegraphics[height=5cm,width=7.5cm]{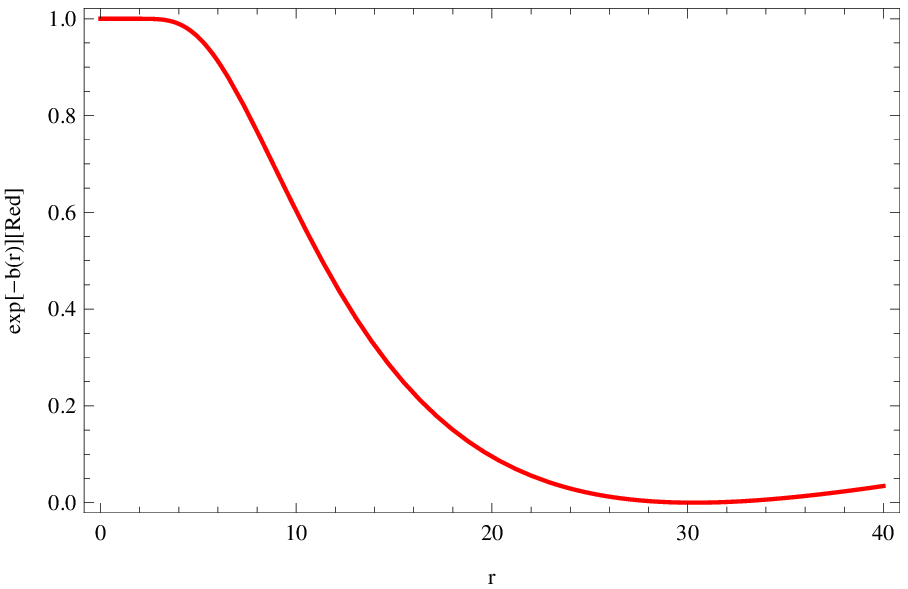}\\
\includegraphics[height=5cm,width=7.5cm]{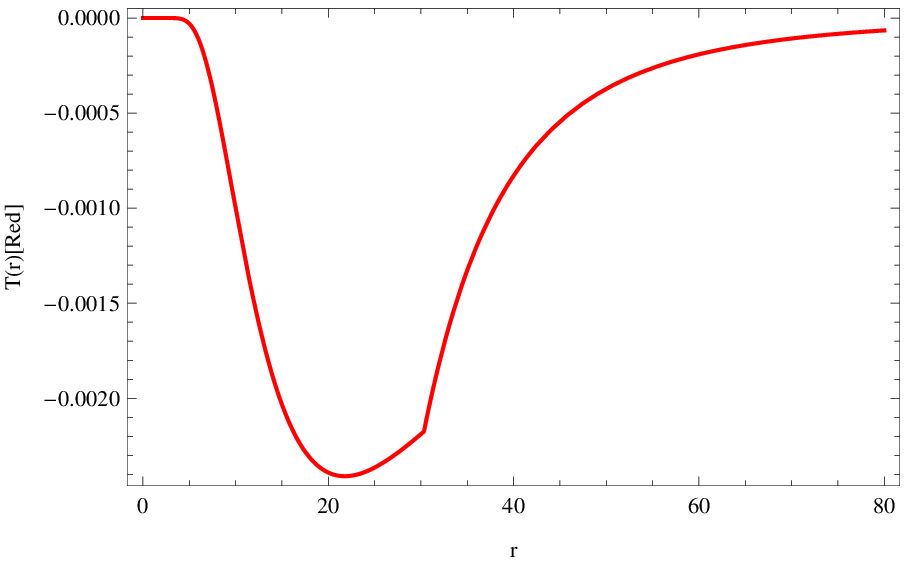}&
\includegraphics[height=5cm,width=7.5cm]{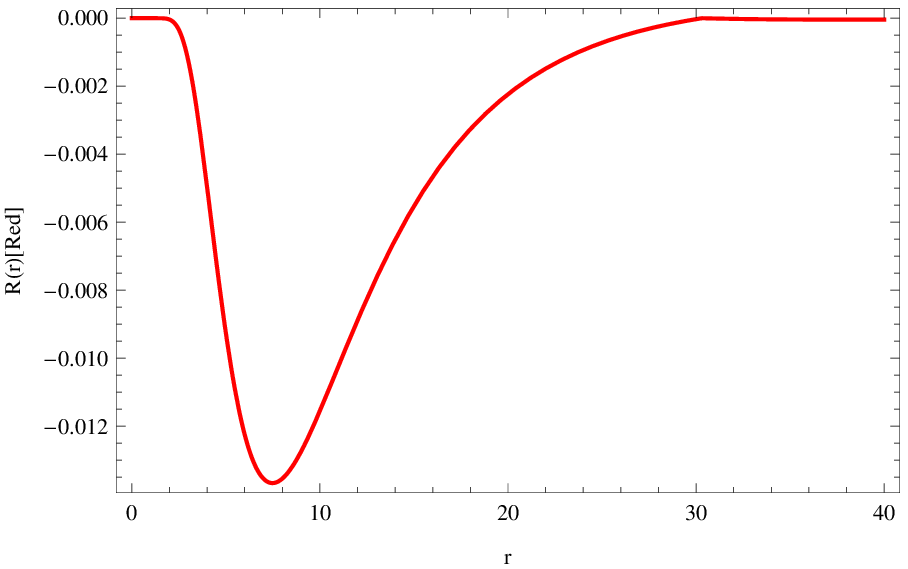}\\
\includegraphics[height=5cm,width=7.5cm]{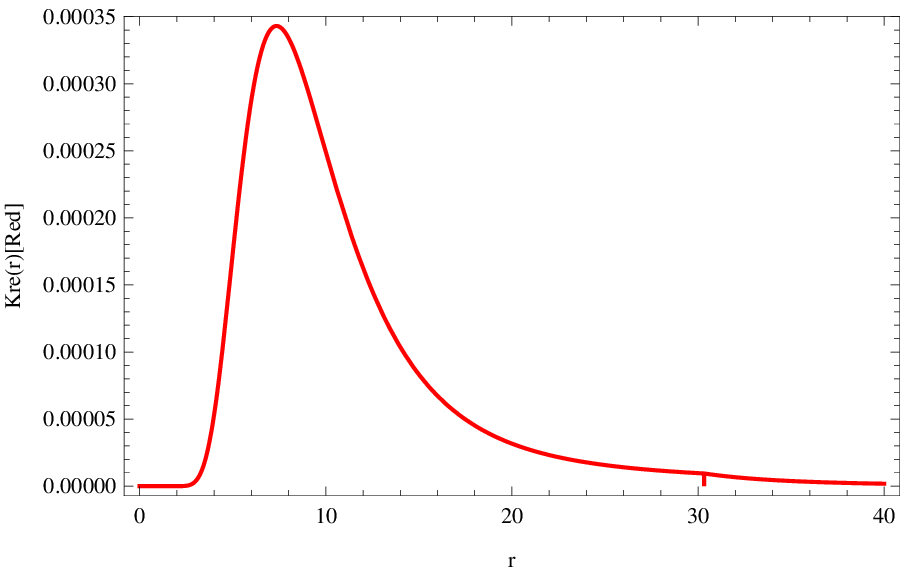}
\end{tabular}
\caption{\scriptsize{Graphical representation of functions $\exp[a(r)],\exp[-b(r)],T(r),R(r)$ and $Kre(r)$, in the extreme case, for the values $\{m_0=\sqrt{e}q/2,q=50\}$.}} 
\label{fig4}
\end{figure}

Let's look again at the metrics and scalar functions, which represent the figure \ref{fig4}. 
We can see that we have only one horizon $r_{eH2}=30.326533058652203$. The torsion scalar $T(r)$, the curvature scalar $R(r)$ and the Kretschmann scalar They are finite in all space-time.
\par
\begin{figure}[h]
\centering
\begin{tabular}{rl}
\includegraphics[height=6cm,width=12cm]{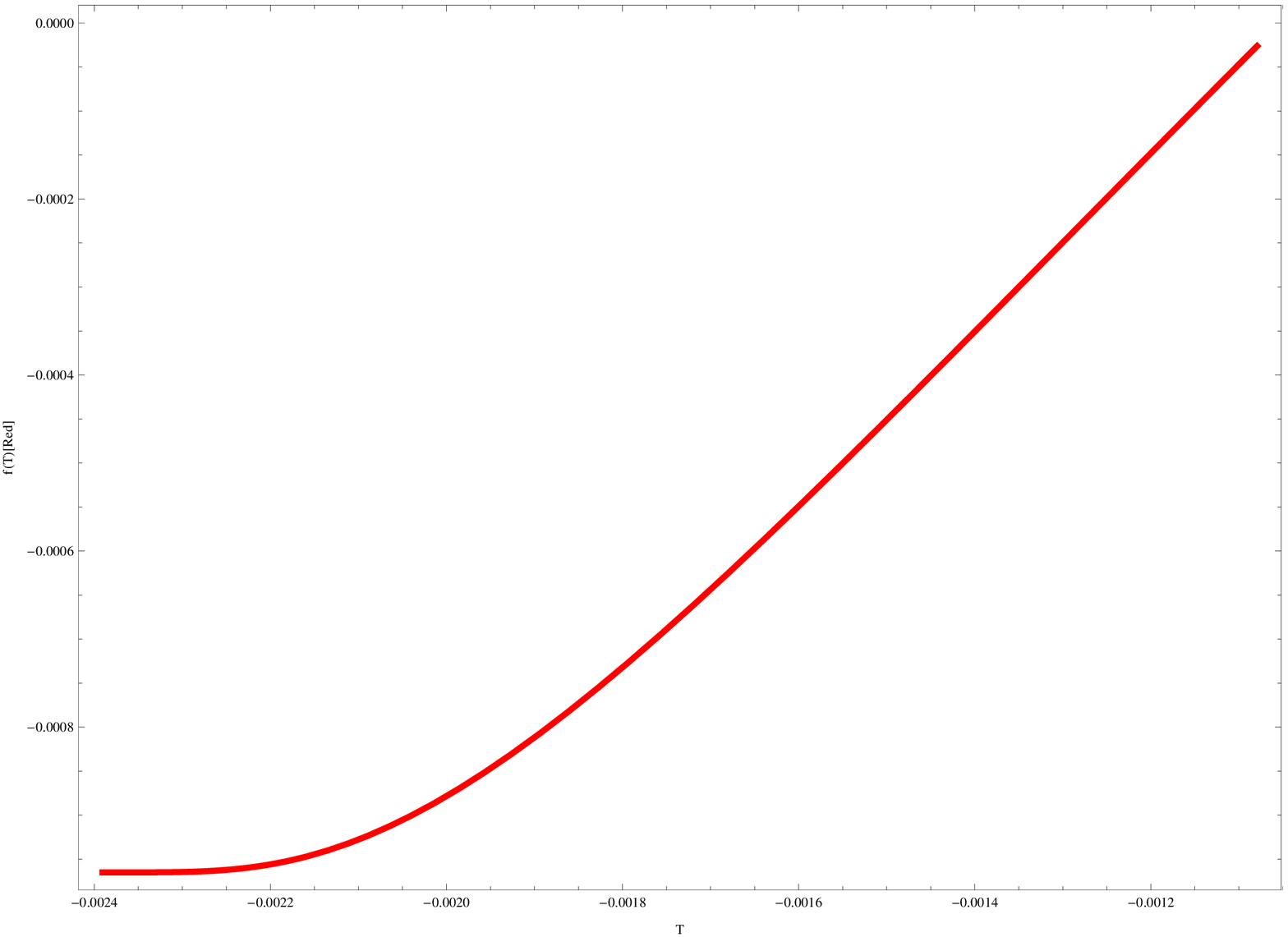}
\end{tabular}
\caption{\scriptsize{Parametric representation to $f(T)\times T$, in the degenerated case for the values $\{m=\sqrt{e}q/2,q=10,r\in[10.3,20]\}$.}} 
\label{fig4-1}
\end{figure}

 
We again calculate numerically the function $f_T(r)$ on (\ref{fT}), using (\ref{a1}) and (\ref{b3}) for this model, then we use the chain rule $df/dT=(df/dr)(dr/dT)$ (integrating to determine $f(r)=\int f_T (dT/dr)dr$) obtain at the end the function $f(r)$, which we can represent using a parametric graph in Figure \ref{fig4-1}. We can see in Figure \ref{fig4-1} that the function $f(T)$ does not depend linearly on $T$, then we are not address the TT here.


\section{Conclusion}\label{sec5}
As well as in GR, by analogy, we establish the equations of motion for the $f(T)$ theory with matter source of non-linear electrodynamics (NED), with spherical symmetry in $4D$, in order to obtain regular black holes solutions. Through the equations of motion we show that for the particular case where $f(T)=T$, we can re-obtain all the solutions previously obtained in GR. Also, through the previous the equations of motion, we show that the unique regular black holes solutions which can be obtained in $f(T)$ theory,  with the non-linear terms in the function and the tetrads chosen, are in conformity with the no-go theorem in \cite{junior} for $b(r)\neq -a(r)$.
\par
We elaborate here a powerful method for obtaining new regular black holes solutions, in the case of $f(T)$ theory with the NED source. Thus, by fixing first the functional form of the metric $a(r)$, we get a new class of black hole (singular), where the scalars, torsion scalar, curvature scalar and $R^{\mu\nu\alpha\beta}R_{\mu\nu\alpha\beta}$, converge to zero at the limit of the origin of the radial coordinate, but diverge for the value $r_{s2}<r_H$. In this case we choose just one model where $M(r)=me^{-q^2/(2mr)}$, with $a(r)=\ln[1-(2M(r)/r)]$ and  $b(r)=2\ln\{[M'(r)-(M(r)/r)-a(r)]/a(r)\}$. In this case the solution is asymptotically flat, possesses an event horizon covering a singular region $r=r_{s2}$. Still study the extreme case, where $m=\sqrt{e}q/2$, resulting in the same structure. Both cases present the same causal structure of the Schwarzschild black hole on GR.
\par 
We also fix the functional form of $b(r)=-\ln[1-(2m(r)/r)]$, obtaining a new class of regular black hole solution. The solution with $m(r)=m_0e^{-q^2/(2m_0r)}$ is asymptotically flat and regular everywhere. They possess a degenerate horizon behaving as the ones corresponding to GR, but here  $b(r)\neq -a(r)$ and a different curvature scalar $R(r)$ given by (\ref{R3}).
\par 
Our perspectives about these new class of charged regular solutions that it is still possible to find another solutions here. We also believe these classes of charged regular solutions must obey the energy conditions for $f(T)$ Gravity \cite{liu}, but breaks specifically  the SEC, as well as in  GR \cite{zaslavskii} (see also \cite{balart2,dymnikova} where the DEC and  WEC are satisfied).  We also think there should be some means to test the requirement that electrically charged solutions do not have the correct weak field limit as in GR \cite{bronnikov}. This should be thoroughly studied on another paper.

\vspace{1cm}

{\bf Acknowledgement}: Ednaldo L. B. Junior thanks CAPES for financial support. Manuel E. Rodrigues  
thanks UFPA, Edital 04/2014 PROPESP, and CNPq, Edital MCTI/CNPQ/Universal 14/2014,  for partial financial support. Mahouton J. S. Houndjo thanks ENS-Natitingou for partial financial support.


%

\end{document}